\documentclass[10pt,a4paper]{article}

\usepackage{amssymb}
\usepackage{amsmath}
\usepackage{bm}
\usepackage{graphicx}
\usepackage{subfigure}
\usepackage{hyperref}
\usepackage{color,soul}

\usepackage{booktabs}
\usepackage{multirow}
\usepackage{array}
\usepackage{enumitem}
\usepackage{cite}
\usepackage[affil-it,auth-lg]{authblk}
\usepackage{eso-pic}

\usepackage{enumitem}

\begin{document}
	
	\title{Secondary component of gravitational-wave signal GW190814 as an anisotropic neutron star}
	\author[1]{Zacharias Roupas}
	\affil[1]{Centre for Theoretical Physics, The British University in Egypt, Sherouk City 11837, Cairo, Egypt} 
	
	\date{\vspace{-5ex}}
	
	\maketitle

\begin{abstract}
	The gravitational-wave signal GW190814 involves a compact object with mass $(2.50-2.67){\rm M}_\odot$ within the so-called low mass gap. As yet, a general consensus on its nature, being a black hole, a neutron star or an exotic star, has not been achieved. We investigate the possibility this compact object to be an anisotropic neutron star. 
	Anisotropies in a neutron star core arise naturally by effects such as superfluidity, hyperons, strong magnetic fields and allow the maximum mass to exceed that of the ideally isotropic stars.
	We consider the Krori-Barua ansatz to model an anisotropic core and constrain the equation of state  with LIGO/Virgo observations GW170817 and GW190814. We find that the GW190814 secondary component can be an anisotropic neutron star compatible with LIGO/Virgo constraints if the radius attains a value in the range $(13.2-14.0)\,{\rm km}$ with the anisotropic core's boundary density  in the range $(3.5-4.0)\cdot 10^{14}{\rm g}/{\rm cm}^3$.
\end{abstract}

\section{Introduction}

The LIGO/Virgo Collaboration announced recently the observation of a merger of a black hole with mass $23.2^{+1.1}_{-1.0} {\rm M}_\odot$ with a compact object with mass $2.59^{+0.08}_{-0.09}{\rm M}_\odot$ \cite{2020ApJ...896L..44A}. The mass of the secondary component lies within the so-called low mass gap \cite{Bailyn_1998,_zel_2010,Belczynski_2012}.
	Theoretical and observational evidence suggest that black holes of mass less than $\sim 5{\rm M}_\odot$ may not be produced by stellar evolution \cite{Bailyn_1998,_zel_2010,Farr_2011}. On the other hand while some candidate equations of state of neutron stars allow for a neutron star maximum mass $M_{\rm max} \sim 3{\rm M}_\odot$ \cite{1996NuPhA.606..508M,PhysRevLett.32.324,_zel_2012,Kiziltan_2013,Alsing_2018}, the relatively small tidal deformabilities measured in gravitational-wave signal GW170817 do not favor such large values of $M_{\rm max}$ but rather suggest it is of the order of $2.5{\rm M}_\odot$
	\cite{PhysRevLett.121.161101,2019PhRvX...9a1001A,2020ApJ...896L..44A}. 
The heaviest neutron star observed to date has a mass of $2.01\pm 0.04{\rm M}_\odot$ \cite{Antoniadis_2013}, while 
there was a recent claim that PSR J0740+6620 may host a
$2.14^{+0.10}_{-0.09} {\rm M}_\odot$ neutron star
\cite{Cromartie_2019}. Therefore the existence and nature of compact objects in the mass regime $\sim {[2.5,5]} {\rm M}_\odot$ are highly uncertain.

These theoretical and observational uncertainties regarding the maximum mass of neutron stars and lower mass of black holes render it challenging to conclude with certainty on the nature of the secondary GW190814 component. Several analyses seem to favor a stellar black hole \cite{2020ApJ...896L..44A,2020arXiv200703799F,2020arXiv200706057T,2020arXiv200709683S}. Other proposals include a
primordial black hole \cite{2020arXiv200615675V,lehmann2020modelindependent,2020arXiv200706481C,2020arXiv200703565J}, a fast pulsar \cite{2020arXiv200702513Z,2020arXiv200614601M,2020arXiv200705526T},
a heavy neutron star with stiff equation of state \cite{2020arXiv200616296T}, 
a neutron star with accretion disk \cite{2020arXiv200700847S}.

Independent of the LIGO/Virgo results, neutron star properties have been reported recently by the Neutron Star Interior Composition Explorer (NICER) team.  In partcular, observations of the isolated millisecond pulsar PSR J0030+0451 indicate a stiffer equation of state \cite{2019ApJ...887L..24M,2019ApJ...887L..21R} than those mostly favored by the LIGO/Virgo collaboration \cite{PhysRevLett.121.161101,2020ApJ...896L..44A}. 

Besides a stiff equation of state, anisotropies in the core of a neutron star \cite{2007ASSL..326.....H} may also allow higher maximum neutron star masses \cite{1975A&A....38...51H}.
The anisotropies inside the star can grow due to superfluidity \cite{1969Natur.224..673B,1970PhRvL..24..775H,2019EPJA...55..167S}, solidification \cite{1971NPhS..231..145A,1972NPhS..236...37C,1973PhRvL..30..999C,1973NPhS..243..130S}, hyperons \cite{1998PhRvC..57..409B}, quarks \cite{1984PhR...107..325B} as well as pion and kaon condensates \cite{1972PhRvL..29..382S,1995PThPh..94..457T}. In addition, nuclear matter in a magnetic field becomes anisotropic, with different pressures in directions along and transverse to the field \cite{2010PhRvC..82f5802F,2019Univ....5..104F}. The electromagnetic energy-momentum tensor is naturally anisotropic. Thus, an anisotropic core is more realistic than an ideally isotropic one.

An anisotropic compact object is subject to a tangential pressure $p_t = p_\theta = p_\phi$ in the angular directions that is different than the radial pressure $p_r \neq p_t$ \cite{1964RSPSA.282..303B,Bowers:1974tgi}. If the anisotropy parameter is positive $\Delta \equiv p_t-p_r>0$ an additional repulsive anisotropic force enables more compact stable configurations to appear in the anisotropic than in the isotropic case \cite{2002PhRvD..65j4011I}. 
The maximum mass of anisotropic compact neutron stars has been estimated to be $M_{\rm max}\sim 4 {\rm M}_\odot$ \cite{1975A&A....38...51H}. Several anisotropic solutions include Refs \cite{PhysRevD.26.1262,PhysRevD.77.027502,Thirukkanesh_2008,2016Ap&SS.361..339S,Maurya_2016,Maurya_2017,2018EPJC...78..673E,2019EPJC...79..885T,2019EPJC...79..853D,2019EPJP..134..600E,2019EPJC...79..138B,Ivanov_2002}.

Here, we will investigate the possibility that the GW190814 secondary component is an anisotropic neutron star.
We will work in a metric ansatz introduced by Krori \& Barua (KB) \cite{1975JPhA....8..508K}. 
Anisotropic compact star models in the KB-spacetime have also been studied in Refs. \cite{PhysRevD.82.044052,2012EPJC...72.2071R,Kalam_2013,Bhar:2014mta,2015Ap&SS.356..309B,Bhar_etaL_2015,2015Ap&SS.359...57A,2016Ap&SS.361....8Z,2016Ap&SS.361..342Z,2016CaJPh..94.1024S,2017Ap&SS.362..237I,2018EPJC...78..307Y,2018IJGMM..1550093S,2019CoTPh..71..599F,2019EPJC...79..919S,2020IJMPA..3550013S,2020MPLA...3550354S,2020arXiv200709797R}.With input the total mass we will calculate the boundary density, radius, and equation of state compatible with LIGO's constraints imposed by gravitational-wave signals GW170817 and GW190814 \cite{2020ApJ...896L..44A}. 

In the next section \ref{sec:GW190814} we impose the LIGO constraints of the neutron star equation of state to an anisotropic neutron star with mass $2.6{\rm M}_\odot$, that is equal to that of the secondary GW190814 component. In the final section we discuss our conclusions. 

\section{Anisotropic star subject to LIGO constraints}\label{sec:GW190814}

Let us first review briefly the KB-ansatz.
We write the spherically symmetric metric in General Relativity in spherical coordinates $(t,r,\theta, \phi)$ as
\begin{equation}
ds^2=-e^{\alpha(r)} \, c^2 dt^2+e^{\beta(r)}dr^2+r^2(d\theta^2+\sin^2\theta d\phi^2)\,,\label{eq:met1}
\end{equation}
for some functions of $r$, $\alpha(r)$ and $\beta(r)$. 
If we denote $\rho$, $p_r$, $p_t$, the mass density, the radial pressure and the tangential pressure, respectively, the spherically symmetric, anisotropic energy momentum tensor is 
\begin{equation}\label{eq:T_enmom}
T_\nu^\mu{} =(\frac{p_t}{c^2}+\rho )u^\mu u_\nu + p_t\delta_a{}^\mu + (p_r - p_t) \xi^\mu \xi_\nu,
\end{equation}
where $u^\mu$ is the four-velocity and $\xi^\mu$ the unit space-like vector in the radial direction. 

Krori \& Barua \cite{1975JPhA....8..508K} introduced the following ansatz for the metric potentials 
\begin{align}\label{eq:pot}
\alpha(x) =a_0 x^2+a_1\,,\quad 
\beta(x) =a_2 x^2\, ,
\end{align}
where however we use the dimensionless variable
\begin{equation}\label{eq:x_dless}
x\equiv \frac{r}{R} \in [0,1].
\end{equation}
We will use the KB-ansatz to model the core of an anisotropic neutron star.
We denote the radius of the core as $r=R$.
Using the characteristic density
\begin{equation}\label{eq:rho_star}
\rho_\star \equiv \frac{c^2}{8\pi G R^2}
\end{equation}
we get the dimensionless variables
\begin{equation}\label{eq:rho+p_dless}
\tilde{\rho} = \frac{\rho}{\rho_\star}\,,\;
\tilde{p}_r = \frac{p_r}{\rho_\star c^2}\,,\;
\tilde{p}_t = \frac{p_t}{\rho_\star c^2}\, .
\end{equation}
Einstein equations give for the dimensionless variables
\begin{align}
\label{eq:rho_a}
\tilde{\rho} &= 
\frac{ e^{-a_2 x^2}}{x^2} \left( e^{a_2 x^2}-1 + 2a_2 x^2 \right)
\,, 
\\
\label{eq:p_r_a}
\tilde{p}_r &= 
\frac{ e^{-a_2 x^2}}{x^2} \left( 1-e^{a_2 x^2}+2a_0 x^2 \right)\,,
\\
\label{eq:p_t_a}
\tilde{p}_t &= 
e^{-a_2x^2}
\left( 2a_0  -a_2 + a_0 (a_0 - a_2) x^2 \right)\, .
\end{align}

Integrating $\mathcal{M}^\prime = 4\pi \rho r^2$ we get for
the mass $\mathcal{M}(r)$ contained within $r$ that
\begin{equation}
\label{eq:mas_a}
\mathcal{M}(x)=M\, C^{-1} x\left( 1-e^{-a_2 x^2}\right) \, .
\end{equation}
where $M$ denotes the total mass of the core and $C$ is the compactness
\begin{equation}
C = \frac{2GM}{Rc^2}.
\end{equation}

We match the KB-solution with the Tolman-Oppenheimer-Volkoff  metric\cite{2020CQGra..37i7001R} at the boundary of the anisotropic core 
\begin{align}
ds^2_{\rm TOV} &= -e^{\alpha_{\rm TOV}(r)} \, c^2 dt^2 + \left( 1 - \frac{2G\mathcal{M}(r)}{rc^2}\right)^{-1} dr^2 + r^2(d\theta^2+\sin^2\theta d\phi^2)\,,\label{eq:TOV}
\\
\alpha_{\rm TOV}(r) &= \frac{2}{c^2}\int_r^\infty d\xi\, 
\left(\frac{G \mathcal{M}(\xi)}{\xi^2} + \frac{4\pi G}{c^2} P(\xi)\, \xi\right)\left( 1 - \frac{2G \mathcal{M}(\xi) }{\xi c^2} \right)^{-1}, \;
r\geq R,
\end{align}
that is assumed to describe the outer layers with a different equation of state $P = P(\rho)$. The boundary pressure $p_R\equiv p_r(R) = P(R)$ is a free parameter. We have
\begin{equation}\label{eq:boundary_cond} 
\beta(r=R)= \ln \left(1- C \right)^{-1}, \quad
\tilde{p}_r(r=R) = \tilde{p}_R.
\end{equation}
For our purposes we only need $a_0$, $a_2$ as is evidenced from Eqs. (\ref{eq:rho_a})-(\ref{eq:p_t_a}). By use of Eqs. (\ref{eq:pot}), (\ref{eq:p_r_a}), (\ref{eq:boundary_cond}) we get  
\begin{equation}
\label{eq:a_param} 
a_0(C) = \frac{1}{2}\frac{C +\tilde{p}_R}{1-C}, 
\quad 
a_2(C)= \ln\left(1- C\right)^{-1}\,.
\end{equation}
These equations along with (\ref{eq:rho_a})-(\ref{eq:p_t_a}) allow us parametrize density and pressure solely with respect to compactness, $\tilde{\rho} = \tilde{\rho}(x;C)$, $\tilde{p} = \tilde{p}(x;C)$, as was firstly shown in \cite{2020arXiv200709797R}. Stars with the same compactness acquire the same core profile and properties. The parameter $a_1$, not required in our analysis, can be determined from the boundary condition $\alpha_{\rm TOV}(r=R) = a_0 + a_1$.

\begin{figure}[!tb]
	\begin{center}
		\includegraphics[scale = 0.5]{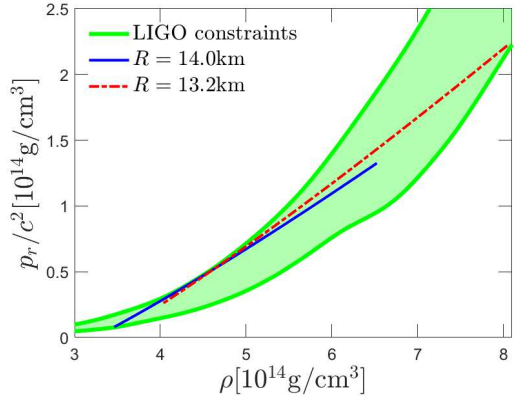}  
		\caption{The radial pressure with respect to the density for the two marginal anisotropic cores with $M = 2.6 {\rm M}_\odot$ and $R = 13.2$ (dash-dotted line), $R=14.0{\rm km}$ (continuous line) compatible with LIGO's constraints (green shaded region) from GW170817, GW190814. }
		\label{fig:eos_LIGO}
	\end{center} 
\end{figure}

The boundary density $\rho_R$ and boundary radial pressure $p_R$ are free parameters which we constrain by LIGO's observations.
These are summarized in Figure 8 of \cite{2020ApJ...896L..44A} and reproduced here as the green shaded region of Figure \ref{fig:eos_LIGO}. We find that an anisotropic neutron core with mass equal to the secondary GW190814 component $M=2.6{\rm M}_\odot$ satisfies the constraints on the equation of state compatible with signals GW170817 and GW190814 given by LIGO only for radius with values in the range $(13.2-14.0){\rm km}$, given in detail in Table \ref{tab:density}. The boundary density is $(3.5-4.0)10^{14}{\rm g}/{\rm cm}^3$ with the lower value corresponding to the bigger radius. In Figure \ref{fig:eos_LIGO} we depict that the two marginal solutions $R=13.2{\rm km}$ and $R = 14.0{\rm km}$ satisfy LIGO's constraints of the equation of state. It seems plausible that the bigger star $R=14.0{\rm km}$ is a more probable candidate, because of the more realistic core's boundary density $\rho_R=3.5\cdot 10^{14}{\rm g}/{\rm cm}^3$.

\begin{small}
	\begin{table}[tbp]
		\begin{center}
			\begin{tabular}{c | c  c  c  c  }
				\toprule
				$R {[{\rm km}]}$ & 
				$\rho_R{[10^{14}{\rm g}/{\rm cm}^3]}$ &
				$p_R/c^2{[10^{14}{\rm g}/{\rm cm}^3]}$ &
				$\frac{dp_r}{d\rho}{[c^2]}$ &
				$\frac{dp_t}{d\rho}{[c^2]}$ 
				\\
				\midrule
				$13.2$
				&
				$4.03$
				&
				$(0.24-0.26)$
				&
				$0.49$
				&
				$0.34$
				\\
				$13.3$
				&
				$3.95$
				&
				$(0.18-0.23)$
				&
				$(0.46-0.48)$
				&
				$(0.32-0.34)$
				\\
				$13.4$
				&
				$3.88$
				&
				$(0.14-0.21)$
				&
				$(0.45-0.47)$
				&
				$(0.31-0.33)$
				\\
				$13.5$
				&
				$3.81$
				&
				$(0.13-0.18)$
				&
				$(0.44-0.46)$
				&
				$(0.30-0.32)$
				\\
				$13.6$
				&
				$3.73$
				&
				$(0.12-0.16)$
				&
				$(0.44-0.45)$
				&
				$(0.30-0.31)$
				\\
				$13.7$
				&
				$3.66$
				&
				$(0.11-0.14)$
				&
				$(0.43-0.44)$
				&
				$0.30$
				\\
				$13.8$
				&
				$3.59$
				&
				$(0.10-0.12)$
				&
				$(0.42-0.43)$
				&
				$(0.29-0.30)$
				\\
				$13.9$
				&
				$3.53$
				&
				$(0.09-0.10)$
				&
				$0.42$
				&
				$0.29$
				\\
				$14.0$
				&
				$3.46$
				&
				$0.08$
				&
				$0.41$
				&
				$0.28$
				\\ 
				\bottomrule
			\end{tabular}
		\end{center} 
		\caption{The radius $R$, boundary density $\rho_R$, boundary radial pressure $p_R$ and slopes of $p_r(\rho)$, $p_t(\rho)$ linear fits, compatible with LIGO's constraints for an anisotropic core with mass $2.6{\rm M}_\odot$.}
		\label{tab:density}
	\end{table}
\end{small}

The acceptable radii correspond to compactness values $C=(0.55-0.58)$. These are not only realistic for a neutron star, but also are lower than $C<0.71$. This is the stability and physical requirement condition for an anisotropic star in KB-spacetime calculated by Roupas \& Nashed \cite{2020arXiv200709797R}. Note that this limit was calculated assuming zero boundary pressure. Nevertheless, it is straightforward to verify that the solutions of Table \ref{tab:density} satisfy the stricter condition that gave the limit $0.71$, that is the strong energy condition \cite{1988CQGra...5.1329K,2017EPJC...77..738I}  
$
\rho c^2 - p_r - 2p_t > 0\,.
$
It is also straightforward to verify that there are satisfied all other conditions of Roupas \& Nashed \cite{2020arXiv200709797R} such as the stability condition for the adiabatic index \cite{1994MNRAS.267..637C,1997PhR...286...53H}
$
\Gamma \equiv \frac{\rho+ p_r}{p_r}\frac{dp_r}{d \rho} > \frac{4}{3}
$
, the causality conditions $v_r < c$, $v_t <c$ and also the condition of stability against cracking  \cite{HERRERA1992206,2007CQGra..24.4631A} 
$0 < v_r{}^2-v_t{}^2 < c^2$.

\section{Conclusions}

We have used the constraints on a neutron star's core equation of state given by LIGO as derived from the gravitational-wave signals GW170817, GW190814 \cite{2020ApJ...896L..44A} in order to investigate if an anisotropic neutron core is compatible with the secondary GW190817 component. We found intriguingly that not only it is compatible, but also that the parameters imposed by these constraints give physical, stable solutions. There is no a-priori reason why this would be the case, especially since LIGO's analysis \cite{2020ApJ...896L..44A} favours a black hole as a candidate for the secondary GW190814 component.

In particular, we conclude that an anisotropic neutron star core with $M=2.6{\rm M}_\odot$ in the KB-ansatz is compatible with LIGO constraints for a radius $R=(13.2-14.0)\,{\rm km}$. The corresponding boundary density for the $R=14.0{\rm km}$ solution is $\rho_R=3.5\cdot 10^{14}{\rm g}/{\rm cm}^3$ that is very close to the nuclear saturation density. For this solution the linear fit of the equation of state gives $p_r \propto 0.41 \rho c^2$ and $p_t \propto 0.28 \rho c^2$.

\bibliography{2020_Roupas_GW_AnNS}
\bibliographystyle{myunsrt}

\end{document}